# A formal proof of the four color theorem

Limin Xiang
Department of Information Science,
Kyushu Sangyo University
3-1 Matsukadai 2-Chome, Higashi-ku, Fukuoka 813-8503, Japan
E-mail: xiang@is.kyusan-u.ac.jp
Tel: +81-92-673-5400, Fax: +81-92-673-5454

**Abstract**: A formal proof has not been found for the four color theorem since 1852 when Francis Guthrie first conjectured the four color theorem. Why? A bad idea, we think, directed people to a rough road. Using a similar method to that for the formal proof of the five color theorem, a formal proof is proposed in this paper of the four color theorem, namely, every planar graph is four-colorable. The formal proof proposed can also be regarded as an algorithm to color a planar graph using four colors so that no two adjacent vertices receive the same color.

**1. Introduction**

Since 1852 when Francis Guthrie first conjectured the four color theorem [1], a formal proof has not been found for the four color theorem. The **four color theorem**, or the **four color map theorem**, states that given any separation of the plane into contiguous regions, called a "map", the regions can be colored using at most four colors so that no two adjacent regions have the same color. Two regions are called *adjacent* only if they share a border segment, not just a point. To precisely state the theorem, it is easiest to rephrase it in **graph theory**. It then states that the vertices of every planar graph can be coloured with at most four colors so that no two adjacent vertices receive the same color, or "every planar graph is four-colorable" for short. Such a graph can be obtained from a map by replacing every region by a vertex, and connecting two vertices by an edge exactly when the two regions share a border segment (not just a corner.) [1].

A computer-assisted proof of the four color theorem was proposed by Kenneth Appel and Wolfgang Haken in 1976. Their proof reduced the infinitude of possible maps to 1,936 reducible configurations (later reduced to 1,476) which had to be checked one by one by computer and took over a thousand hours [1]. However, because part of the proof consisted of an exhaustive analysis of many discrete cases by a computer, some mathematicians do not accept it [2].

A **graph** $G$ consists of a non-empty finite set $V$ of vertices and a finite set $E$ of edges on $V$, i.e., any edge of $E$ connects two vertices of $V$. If vertex $v$ is shared by $k$ edges, $k$ is called the **degree** of vertex $v$, and denoted by $d(v) = k$.

A **planar graph** $G$ is a Graph that may be embedded in the plane without intersecting edges.

A graph $G$ is said to be $n$-**colorable**, denoted by $c(G) = n$, if it's possible to assign one of $n$ colors to each vertex in such a way that no two connected vertices have the same color.

More than 150 years passed, but a formal proof has not been found for the four color theorem. Why? The following idea [1], we think, directed people to a bumpy road.

> If there is a maximal planar graph requiring 5 colors, then there is a *minimal* such graph, where removing any vertex makes it four-colorable. Call this graph $G$. $G$ cannot have a vertex of degree 3 or less, because if $d(v) \leq 3$, we can remove $v$ from $G$, four-color the smaller graph, then add back $v$ and extend the four-coloring to it by choosing a color different from its neighbors.
>
> Kempe also showed correctly that $G$ can have no vertex of degree 4.

The minimal graph $G$ above does not exist, if the four color theorem holds. Any research on the inexistent minimal graph $G$ should be not only very difficult and inefficient, but also too hard to understand. Kenneth Appel and Wolfgang Haken's computer-assisted proof is just such an example.

In fact, the four color theorem holds, so we may prove it directly. Using a similar method to that for the formal proof of the five color theorem [3], a formal proof of the four color theorem is proposed in this paper. In Section 2, some notations are introduced, and the formal proof of the four color theorem is given in Section 3. A hand-checked case flow chart is shown in Section 4 for the proof, which can be regarded as an algorithm to color a planar graph using four colors so that no two adjacent vertices receive the same color. Section 5 concludes the paper.

**2. Notations**

Given a graph $G$ and a vertex $v$ in $G$, $G-v$ is the graph removing $v$ and all its shared edges from $G$. The reverse of $G-v$ is $(G-v)+v$, and $(G-v)+v=G$.

$G(i,j)$ is a subgraph of $G$ consisting of the vertices that are colored with colors $i$ and $j$ only, and edges connecting two of them. $G^c(i,j,v)$ is the connected component of $G(i,j)$ containing vertex $v$.

A path in $G(i,j)$, called a [Kempe chain]() and denoted by $Ch(i,j,u,v)$, joining vertices $u$ and $v$, that is a sequence of edges and vertices painted only with colors $i$ and $j$.

$v$ is a vertex of graph $G$, which is denoted by $v \in G$, and on the contrary $v$ is not a vertex of graph $G$, which is denoted by $v \notin G$.

A circle is a closed path. When vertex $w \notin Ch(i,j,u,v)$, $Ch(i,j,u,v)$ together with $w$ as well as its two edges connected to $u$ and $v$ forms a **Kempe circle**, which is denoted by $Ch(i,j,u,v)+w$.

Suppose $n(v)$, $n(e)$, and $n(f)$ are the number of vertices, edges, and faces in a planar graph. Since each edge is shared by two faces and each face is bounded by three edges at least, $2n(e) \geq 3n(f)$ which together with **Euler's formula** $n(v)-n(e)+n(f)=2$ can be used to show $6n(v)-2n(e) \geq 12$. Now, if $n(d_i)$ is the number of vertices of degree $d_i$ and $D$ is the maximum degree, with **Euler's Theorem** $2n(e) = \sum_{i=1}^{D} i \times n(d_i)$,

$$6n(v) - 2n(e) = 6\sum_{i=1}^{D} n(d_i) - \sum_{i=1}^{D} i \times n(d_i)$$
$$= \sum_{i=1}^{D}(6-i)n(d_i) \geq 12.$$

But since $12 > 0$ and $6-i \leq 0$ for all $i \geq 6$, this demonstrates that there is at least one vertex of degree 5 or less in a planar graph [1]. Thus, the following lemma holds.

**Lemma 1.** For any planar graph $G_n$ with $n$ ($n \geq 6$) vertices, there are vertices $v_n$, $v_{n-1}$, ..., $v_6$ such that $d(v_i) \leq 5$ and $G_{i-1} = G_i - v_i$ are also planar graphs for $i$ from $n$ down to $6$.

Thus, the formal proof of the four color theorem can be given in the following section.

**3. The proof**

**Theorem 1(The Four color Theorem)** Every planar graph is four-colorable.

**Proof.** Let the planar graph be with $n$ vertices, where $n \geq 1$, and denoted by $G_n$. There are 3 cases (Case.1 – Case.3) to discuss.

**Case.1**: When $1 \leq n \leq 4$, the result holds obviously.

**Case.2**: When $n=5$, the maximal planar graph with $5$ vertices is the full graph deleting an edge, i.e., the planar graph with $5$ vertices and $9$ edges, which is denoted by $G_5'$. Any $G_5$ is a subgraph of $G_5'$, and $c(G_5) \leq c(G_5')$. Since $c(G_5')=4$ (see, for example, Fig. 1(a), ) $c(G_5)=4$, i.e., the result holds.

**Case.3**: When $n \geq 6$, by Lemma 1, there are vertices $v_n$, $v_{n-1}$, ..., $v_6$ such that $d(v_i) \leq 5$ and $G_{i-1} = G_i - v_i$ are also planar graphs for $i$ from $n$ down to $6$.

It will be shown that $c(G_i) = c(G_{i-1} + v_i) = 4$ for $i$ from $6$ up to $n$ in the following.

For $i$ from $6$ up to $n$, since $c(G_5)=4$ by Case.2, let $c(G_{i-1})=4$, and

$$C(v_i) = \{c(u) \mid c(u) \text{ is the color of vertex } u$$

in $G_{i-1}$, and $u$ is adjacent to $v_i$ in $G_i\}$

Note that $C(v_i) \leq 4$ and $C(v_i) \leq d(v_i) \leq 5$. There are 3 cases (Case.3.1 - Case.3.3) to discuss.

**Case.3.1**: $|C(v_i)| \leq 3$, let $v_i$ be colored with the 4$^{th}$ color, and $c(G_i) = 4$. The result holds.

**Case.3.2**: $|C(v_i)| = 4$ and $d(v_i) = 4$, let $v_1, v_2, v_3, v_4$ be adjacent to $v_i$ in clockwise order, and $c(v_j) = j$ for $1 \leq j \leq 4$ without loss of generality. Consider the subgraph $G_i(1,3)$ of $G_i$, there are 2 cases (Case.3.2.1, Case.3.2.2) to discuss.

**Case.3.2.1**: $v_1 \notin G_i^c(1,3,v_3)$ (see, for example, Fig. 1(b),) we can reverse the coloration on $G_i^c(1,3,v_3)$, thus assigning color number 3 to $v_i$ and completing the task.

**Case.3.2.2**: $v_1 \in G_i^c(1,3,v_3)$, we can find a Kempe chain $Ch(1,3,v_1,v_3)$ in $G_i(1,3)$ (see, for example, Fig. 1(c).) Kempe Circle $Ch(1,3,v_1,v_3) + v_i$ separates $G_i^c(2,4,v_2)$ of $G_i(2,4)$ from $G_i^c(2,4,v_4)$, we can reverse the coloration on $G_i^c(2,4,v_2)$, thus assigning color number 2 to $v_i$ and completing the task.

Thus, the result holds in Case.3.2.

**Case.3.3**: $|C(v_i)| = 4$ and $d(v_i) = 5$, let $v_1, v_2, v_3, v_4, v_5$ be adjacent to $v_i$ in clockwise order. Only two of the five vertices are painted with the same color, and the two vertices with the same color are neighbor or isolate in cyclic order.

Therefore, there are 2 cases (Case.3.3.1, Case.3.3.2) to discuss.

**Case.3.3.1**: The two vertices with the same color are neighbor in cyclic order, without loss of generality, let $c(v_j) = j$ for $1 \leq j \leq 4$ and $c(v_5) = 1$. Consider the subgraph $G_i(1,3)$ of $G_i$, there are 2 cases (Case.3.3.1.1, Case.3.3.1.2) to discuss.

**Case.3.3.1.1**: $v_3 \notin G_i^c(1,3,v_1)$ and $v_3 \notin G_i^c(1,3,v_5)$ (see, for example, Fig. 1(d),) we can reverse the coloration on $G_i^c(1,3,v_3)$, thus assigning color number 3 to $v_i$ and completing the task.

**Case.3.3.1.2**: $v_3 \in G_i^c(1,3,v_1)$ or $v_3 \in G_i^c(1,3,v_5)$, we can find a Kempe chain $Ch(1,3,v_1,v_3)$ or $Ch(1,3,v_5,v_3)$ in $G_i(1,3)$ (see, for example, Fig. 1(e).) Kempe Circle $Ch(1,3,v_1,v_3) + v_i$ or $Ch(1,3,v_5,v_3) + v_i$ separates $G_i^c(2,4,v_2)$ of $G_i(2,4)$ from $G_i^c(2,4,v_4)$, we can reverse the coloration on $G_i^c(2,4,v_2)$, thus assigning color number 2 to $v_i$ and completing the task.

Thus, the result holds in Case.3.3.1.

**Case.3.3.2**: The two vertices with the same color are isolate in cyclic order, without loss of generality, let $c(v_1) = c(v_3) = 1$, $c(v_2) = 2$, $c(v_4) = 3$, and $c(v_5) = 4$. Consider the subgraph $G_i(2,4)$ of $G_i$, there are 2 cases (Case.3.3.2.1, Case.3.3.2.2) to discuss.

**Case.3.3.2.1**: $v_2 \notin G_i^c(2,4,v_5)$ (see, for example, Fig. 1(f),) we can reverse the coloration on $G_i^c(2,4,v_5)$, thus assigning color number 4 to $v_i$ and completing the task.

**Case.3.3.2.2**: $v_2 \in G_i^c(2,4,v_5)$. There are 2 cases (Case.3.3.2.2.1, Case.3.3.2.2.2) to discuss.

**Case.3.3.2.2.1**: $v_2 \notin G_i^c(2,3,v_4)$ (see, for example, Fig. 1(g),) we can reverse the coloration on $G_i^c(2,3,v_4)$, thus assigning color number 3 to $v_i$ and completing the task.

**Case.3.3.2.2.2**: $v_2 \in G_i^c(2,3,v_4)$, we can find a Kempe chain $Ch(2,3,v_2,v_4)$ in $G_i(2,3)$ (see, for example, Fig.1(h).) Kempe circle $Ch(2,3,v_2,v_4) + v_i$ separates $G_i^c(1,4,v_3)$ of $G_i(1,4)$ from $G_i^c(1,4,v_5)$, we can reverse the coloration on $G_i^c(1,4,v_3)$. Then (see, for example, Fig. 1(i),) we can find a Kempe chain $Ch(2,4,v_2,v_5)$ in $G_i(2,4)$. Kempe circle $Ch(2,4,v_2,v_5)+v_i$ separates $G_i^c(1,3,v_1)$ of $G_i(1,3)$ from $G_i^c(1,3,v_4)$, we can reverse the coloration on $G_i^c(1,3,v_1)$, thus assigning color number 1 to $v_i$ and completing the task (see, for example, Fig. 1(j).)

Thus, $c(G_i) = c(G_{i-1} + v_i) = 4$, since $c(G_{i-1}) = 4$.

By Mathematical induction, the result holds for all cases.

Therefore, the proof is completed.

**4. The flow chart**

A hand-checked case flow chart is shown in Fig. 2 for the proof in Section 3.

**5. Conclusions**

The hand-checked case flow chart shown in Section 4 can be regarded as an algorithm to color a planar graph using four colors so that no two adjacent vertices receive the same color.

Since the four color theorem holds, it can be proved directly. Using a similar method to that for the formal proof of the five color theorem [3], a formal proof of the four color theorem was proposed in this paper.

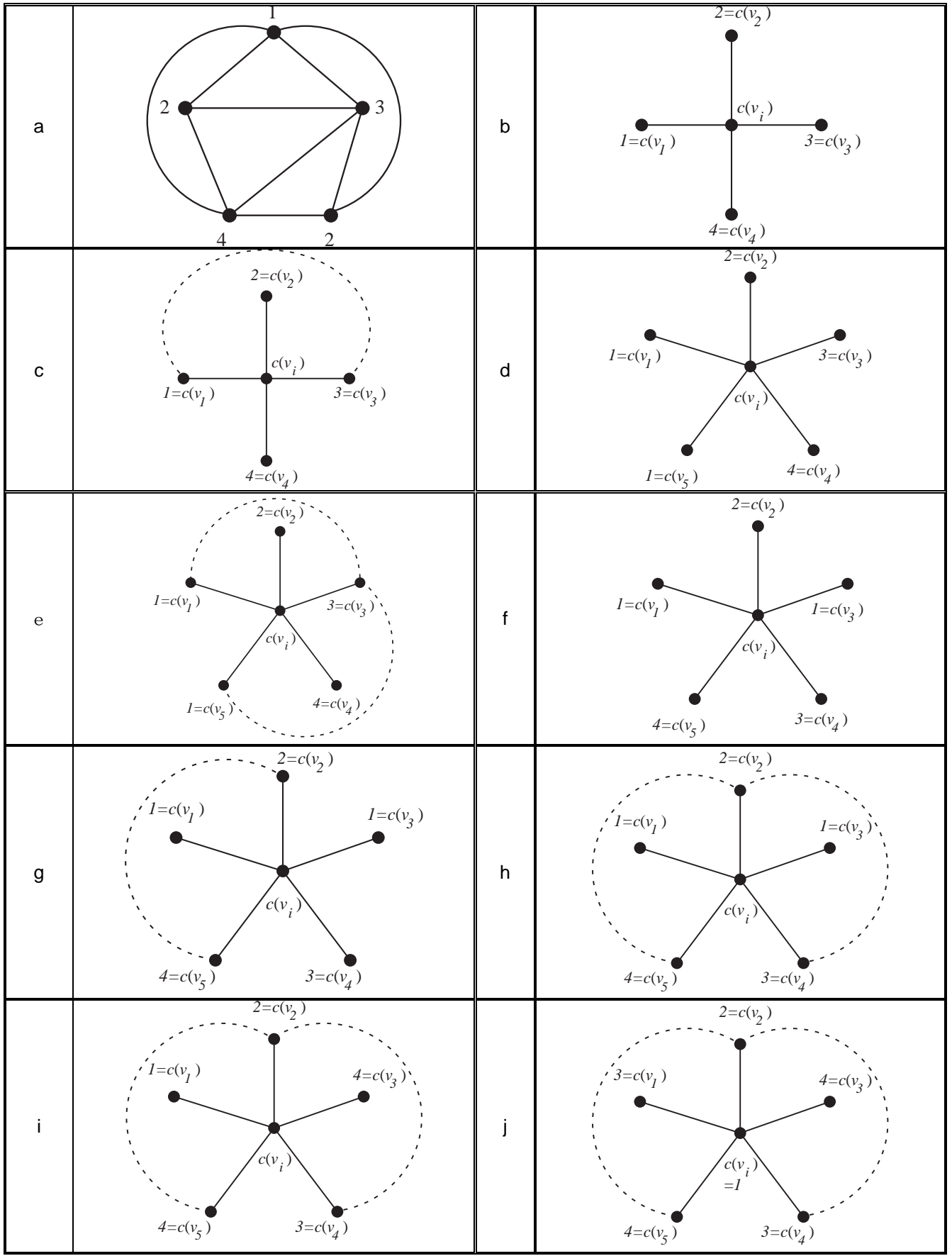

**Fig.1**. For cases of the prof

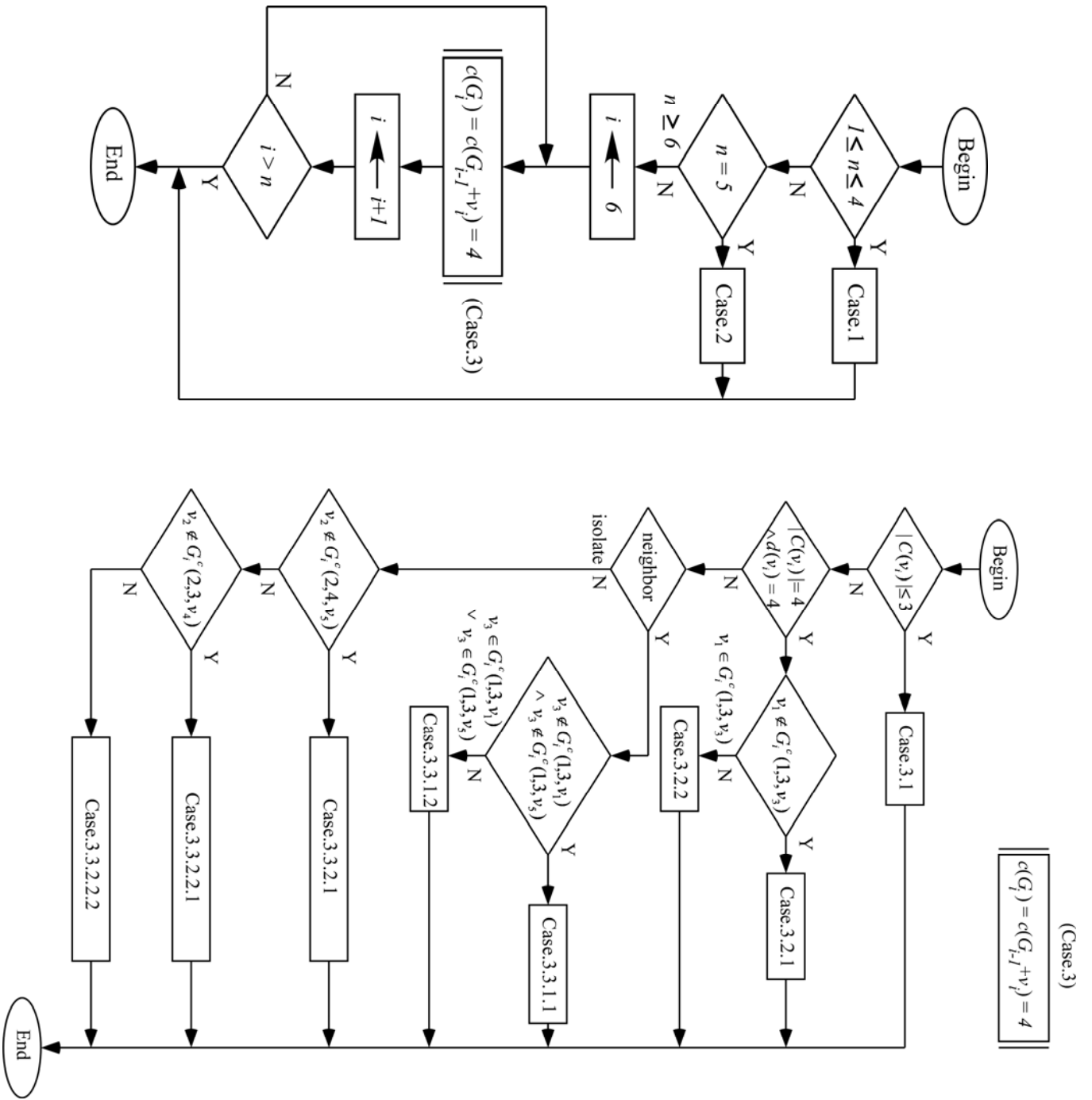

**Fig. 2. The flow chart for the proof**